\documentclass[a4paper]{jpconf}
\usepackage{graphicx}
\newcommand\bk{{\bf k}}
\newcommand\bpi{{\mbox{\boldmath$\pi$}}}
\begin{document}
\title{Magnetic aspects of QCD at finite density and temperature}

\author{Toshitaka Tatsumi}

\address{Department of Physics, Kyoto University, Kyoto 606-8502, Japan}

\ead{tatsumi@ruby.scphys.kyoto-u.ac.jp}

\begin{abstract}
Some magnetic aspects of QCD are discussed at finite density and temperature. Possibility of spontaneous magnetization is studied within Landau Fermi-liquid theory, and the important roles of the screening effects for gluon propagation are elucidated. Static screening for the longitudinal gluons improves the infrared singularities, while the transverse gluons receive only dynamic screening. The latter property gives rise to a novel non-Fermi-liquid behaviour for the magnetic susceptibility.  Appearance of a density-wave state is also discussed in relation to chiral transition, where pseudoscalar condensate as well as scalar one takes a spatially non-uniform form in a chirally invariant way. Accordingly magnetization of quark matter oscillates like spin density wave. A hadron-quark continuity is suggested in this aspect, remembering pion condensation in hadronic phase.  
\end{abstract}

\section{Introduction}
The phase diagram of QCD has been explored by many authors on temperature-density plane \cite{wam}. 
Here we concentrate upon some magnetic phases at moderate density and relatively low temperature. Considering the spin degrees of freedom, we may expect pion condensation (PIC) in hadronic phase, ferromagnetic (FM) or spin density wave (SDW) phase in quark matter. 
\begin{figure}[h]
\begin{center}
\includegraphics[width=0.4\textwidth]{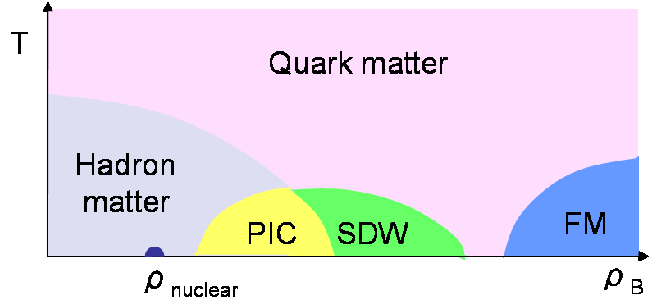}
\caption{Schematic view of the magnetic phase diagram at finite density and temperature.}
\end{center}
\end{figure}
Classical pion field of a shape $\propto \sin(k_cz)$ develops in the PIC phase, followed by anti-ferromagnetic ordering of nucleon spin with a period of $2\pi/k_c$ ({\it alternating-layer-spin} [ALS] structure) \cite{pio}. 

We shall see that a kind of density-wave (call it dual-chiral-density-wave (DCDW)) may appear near the phase boundary of chiral phase transition \cite{dcdw}. The non-uniform pseudoscalar condensate as well as scalar one behaves like a spiral in the chiral space, and accordingly magnetization of quark matter also oscillates in space like SDW. This phase may then be characterized by local ferromagnetism and global anti-ferromagnetism.

Finally it should be interesting to see a possibility of uniform magnetization  at some domain in the density-temperature plane \cite{tat00,tat082,tat083}. If such phase is realized inside compact stars, it gives rise to a direct consequence through their magnetic evolution. The origin of the strong magnetic field is a long-standing problem since the first discovery of pulsars. Recent discovery of magnetars with huge magnetic field ($10^{15}$G) seems to revive this issue \cite{mag}.

Theoretically it should be interesting to see that these phases are triggered by the instability of particle-hole excitations and they are specified by the color-singlet order parameters. So these phases survive in the large $N_c$ limit. Moreover, we will see some $N_f$ dependence also appears.

\section{Ferromagnetism and magnetic susceptibility}

In ref.~\cite{tat00} we have studied a possible ferromagnetic transition in QCD due to the Bloch mechanism, stimulated by the discovery of magnetars. We have evaluated the energy difference between spin polarized and non-polarized phases by using the one-gluon-exchange (OGE) interaction.
 Since quark matter is color neutral as a whole, the exchange energy gives the leading-order contribution. Using the relation,  $\langle\lambda_i\rangle_{ab}\langle\lambda_i\rangle_{ba}=1/2-1/(2N_c)\delta_{ab}$, we can see that it repulsively works for any quark pair.  
Then two particles with the same spin can avoid the repulsive interaction due to the Pauli principle to favor spin polarization \cite{blo}. Our result has shown a weakly first-order phase transition around the nuclear density. 

Assuming such phase inside compact stars, we have roughly estimated the magnitude of magnetic field to find $10^{15-17}$G. Since it is consistent with observation of magnetars, QCD may give a microscopic origin of the magnetic field \cite{mag}.  

To get more insight into the ferromagnetic properties of quark matter we have recently studied magnetic susceptibility within Landau Fermi-liquid theory \cite{tat082,tat083}. Magnetic susceptibility is then a measure to the response of quark matter to applying magnetic field $B$,
$ 
\chi_M=\partial M /\partial B|_{N,T,B=0},
$
with magnetisation $ M =\langle\bar q\sigma_{12} q\rangle$. Then divergence of $\chi_M$ is a signal of ferromagnetic phase transition.
Within the Fermi liquid theory it can be expressed in terms of the Landau-Migdal parameters:
\begin{equation}
\chi_M=\left(\frac{g_D\mu_q}{2}\right)^2/\left(\frac{\pi^2}{N_ck_FE_F}-\frac{1}{3}f_1^s+\bar{f}^a\right),
\label{chim}
\end{equation}
where $f_1^s,\bar{f}^a$ are spin-independent and -dependent Landau-Migdal parameters, respectively. 

These parameters usually includes infrared (IR) divergences in gauge theories QCD/QED, so that it is essential to take into account the screening effect to improve them. The HDL resummation can be achieved by using the quark polarization function; longitudinal gluons are statically screened in terms of the Debye mass, while transverse gluons are only dynamically screened due to the Landau damping. Thus Debye screening surely improves the IR divergence for longitudinal gluons, while there still remain the IR divergences in the Landau-Migdal parameters coming from transverse gluons. At $T=0$ these divergences cancel each other in Eq.~(\ref{chim}) to give a meaningful result. We shall see an interesting effect caused by the dynamic screening in section 2.2. 

\subsection{Magnetic transition at $T=0$}

The magnetic susceptibility at $T=0$ is finally given as
\begin{equation}
\left(\chi_M/ \chi_{\rm Pauli}\right)^{-1}=1
-\frac{C_fg^2N_c\mu}{12\pi^2E_F^2k_F}\Big[m(2E_F+m)
-\frac{1}{2}(E_F^2+4E_Fm-2m^2)
\kappa\ln\frac{2}{\kappa} \Big],
\label{scr}
\end{equation}
with $\kappa=m_D^2/2k_F^2$, where the Debye screening mass can be written as 
$m_D^2=\sum_{\rm flavors}g^2/2\pi^2k_{F,i}E_{F,i}$. We present some numerical results in Fig.~2.
\begin{figure}[h]
\begin{center}
\includegraphics[width=0.4\textwidth]{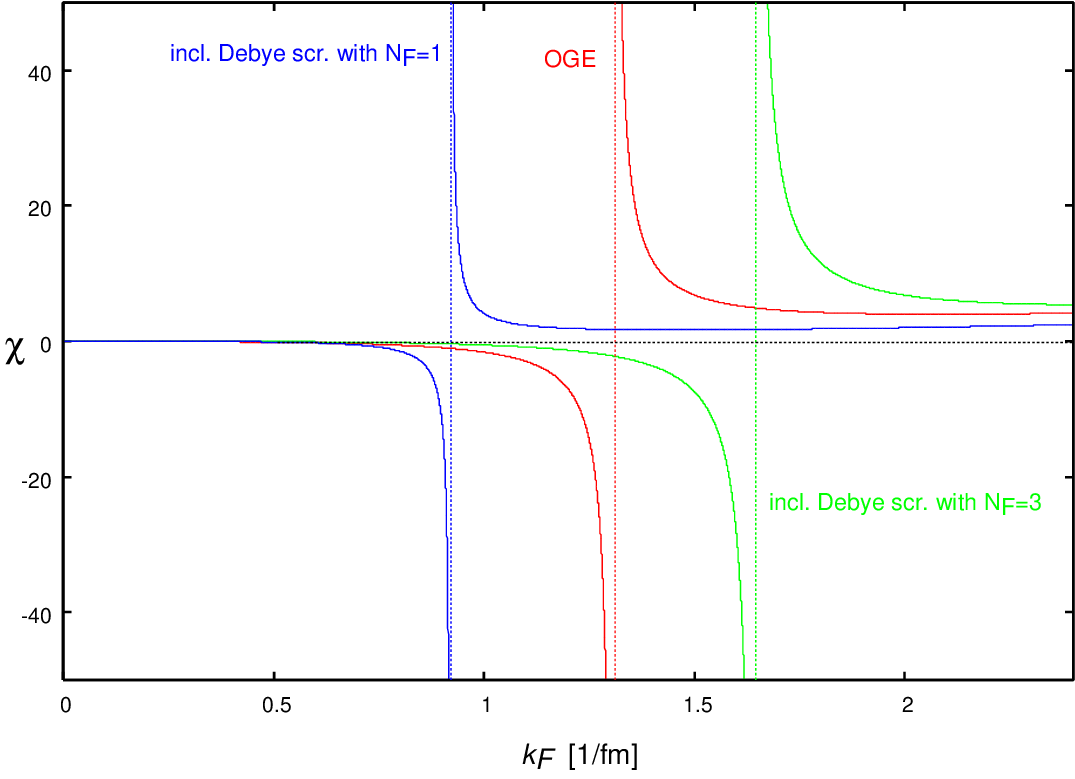}
\caption{Magnetic susceptibility at $T=0$.  The solid curve shows the result
 with the simple OGE without screening, while the dashed and dash-dotted ones 
 shows the screening effect with $N_f=1$ (only $s$ quark)and $N_f=2+1$
 ($u,d,s$ quarks), respectively.}
\end{center}
\end{figure}
We can see that magnetic susceptibility diverges around the nuclear density for pure OGE interaction; quark matter is in ferromagnetic phase below the critical density. When the screening effects are taken into account, the curve is shifted in two ways, depending on the number of flavors; it shifted to lower densities for $N_f=1$, while to higher densities for $N_f=3$. Thus the screening effects favors the ferromagnetic phase for $N_f=3$, different from the case of $N_f=1$. This is in contrast with the usual argument for electron gas, where the correlation effect is always disfavors the magnetic transition \cite{bru57}. Such behavior can be easily understood by looking at $\kappa\ln\kappa$ term in $\chi_M$ (Eq.~(\ref{scr})).

\subsection{Thermal effects and non-Fermi-liquid behavior}

Our framework can be easily extended to finite temperature case \cite{tat083}. We, hereafter, consider the low temperature case ($T/\mu\ll 1$), but the usual low-T expansion cannot be applied since quark energy exhibits an anomalous behavior near the Fermi surface. The Schwinger-Dyson equation gives the quark self-energy; the one-loop result can be written as
\begin{equation}
{\rm Re}\Sigma_+(\omega)\simeq{\rm Re}\Sigma_+(\mu)
-\frac{C_fg^2v_F}{12\pi^2}(\omega-\mu)\ln \frac{m_D}{|\omega-\mu|}+\Delta^{\rm reg}(\omega-\mu) 
\end{equation} 
near the Fermi surface, with $v_F$ being the Fermi velocity and $C_f=(N_c^2-1)/2N_c$ \cite{man}. The second term is provided by the transverse gluons and shows an anomalous behavior near the Fermi surface, while the contribution of the longitudinal gluons is summarized in the regular function $\Delta^{\rm reg}(\omega-\mu)$ of $O(g^2)$; the renormalization factor $z_+(k)$ is then given by the equation,
 $z_+(k)=(1-\partial{\rm Re}\Sigma_+(\omega)/\partial\omega|_{\omega=\epsilon_\bk})^{-1}$, and we find a leading order contribution as  
\begin{equation}
z_+(k)^{-1}\sim -\frac{C_fg^2v_F}{12\pi^2}\ln|\epsilon_\bk-\mu|,
\end{equation}
which exhibits a logarithmic divergence at the Fermi surface. Thus quark matter behaves as the {\it marginal} Fermi liquid \cite{smi} 
\footnote{Note that the effective coupling is now not $g^2$ but $g^2_{\rm eff}=g^2v_F/4\pi$ and the renormalization-group argument shows that it is relevant and infrared free, i.e. $g^2_{\rm eff}$ is getting weaker as $\omega$ approaches $\mu$ \cite{sch}. Thus one-loop result well approximates the solution of the Schwinger-Dyson equation even for rather large $\alpha_s$ \cite{deu}.}
. 

The temperature dependent term in $\chi_M$ is given by 
\begin{eqnarray}
\delta\chi_M^{-1}&=&\chi_{\rm Pauli}^{-1}\Big[ \frac{\pi^2}{6k_F^4}\left(2E_F^2-m^2+\frac{m^4}{E_F^2}\right)T^2 \nonumber\\
&&+\frac{C_fg^2v_F}{72k_F^4E_F^2} \left(2k_F^4+k_F^2m^2+m^4\right)T^2\ln\left(\frac{m_D}{T}\right)\Big]+O(g^2T^2).
\label{delT}
\end{eqnarray}
We can see that there appears $T^2\ln T$ term beside the usual $T^2$ one. This is a novel non-Fermi-liquid effect \cite{tat083}. It should be interesting to compare this result with other typical non-Fermi-liquid effect in the specific heat \cite{hol} or the gap function in color superconductivity \cite{son}.  

In Fig.~3 we plot the magnetic susceptibility at finite temperature, by changing $T$ from $0$ to $60$ MeV. One may see that the ferromagnetic region shrinks as increasing $T$, and there shows no longer divergence at $T=60$ MeV.
Finally the magnetic phase digram is presented on temperature-density plane (Fig.~4).
\begin{figure}[h]
\begin{minipage}{0.39\textwidth}
\begin{center}
\vspace{0.2cm}
\includegraphics[width=5cm]{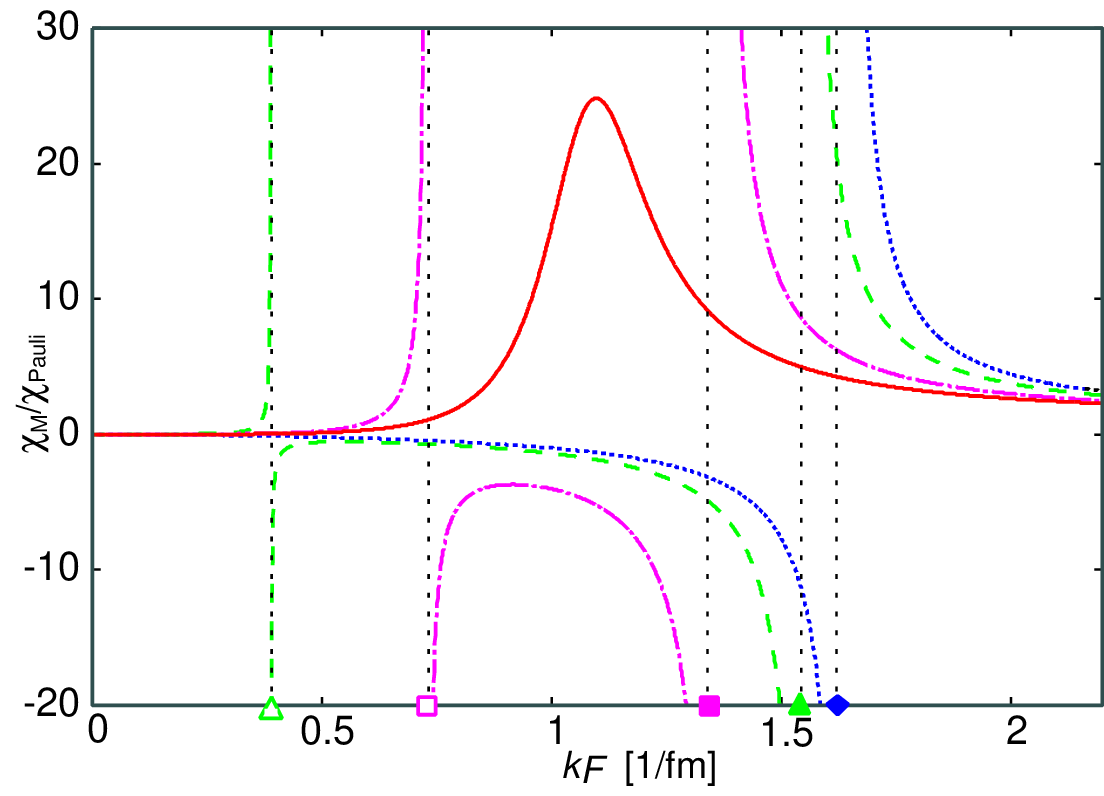}
\vspace{0.3cm}
\caption{Magnetic susceptibility at $T\neq 0$. The dotted, dashed, dash-dotted and solid curves show the results at $T=0, 30, 50$ and $60$ MeV, respectively.}
\label{magsusept}
\end{center}
\end{minipage}
\hspace{\fill}
\begin{minipage}{0.5\textwidth}
\begin{center}
\includegraphics[width=6cm]{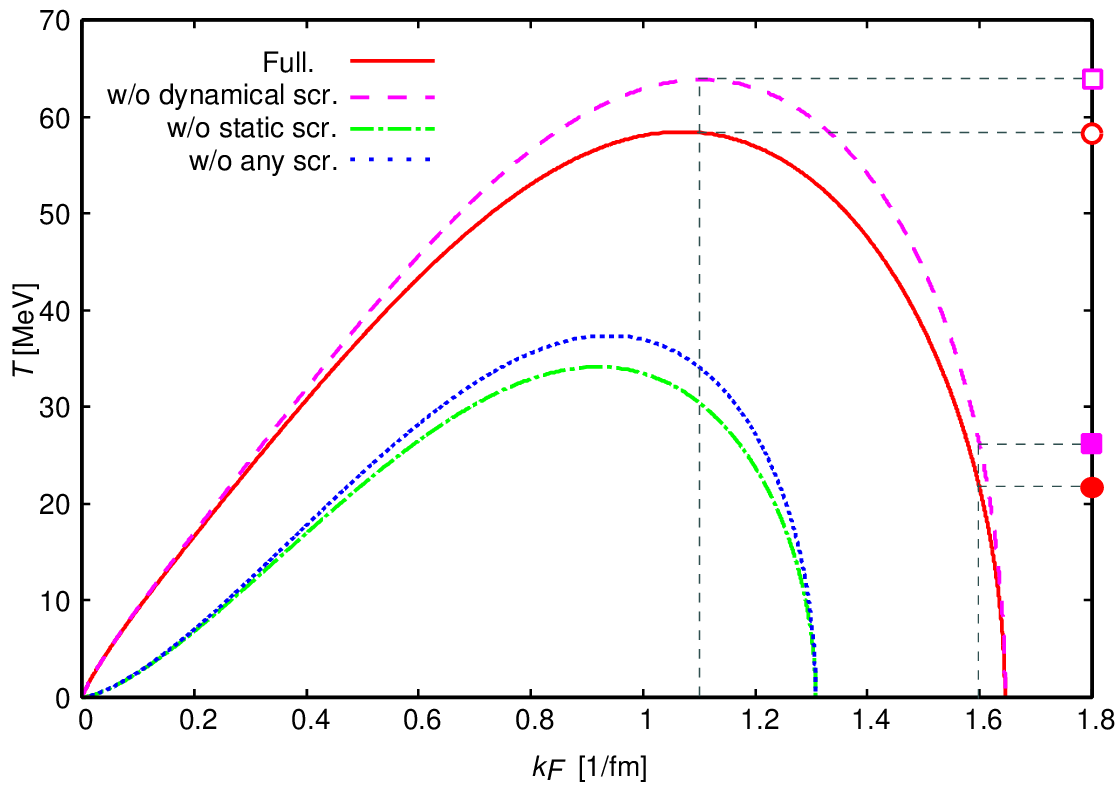}
\caption{Magnetic phase diagram on the density-temperature plane. The
 solid, dashed, dash-dotted, dotted curves show the results for the full
 expression, the one without the $T^2 \ln T$ term,
 without the $\kappa \ln \kappa$ term, and without the $T^2 \ln T$ and
 $\kappa \ln \kappa$ terms. }
\label{phdiagram}
\end{center}
\end{minipage}
\end{figure}

\section{Magnetism and chiral symmetry}

We have discussed ferromagnetic transition by using OGE interaction so far, while non-perturbative effects should remain at moderate densities. Moreover, restoration of chiral symmetry is also important there. So we must take into account these effects for more realistic description of magnetic aspects of QCD. Taking 2 flavor NJL model as an effective model of QCD, respecting chiral symmetry, we can see some inherent properties to produce magnetism. The pseudoscalar interaction $G(\bar\psi i\gamma_5\mbox{\boldmath$\tau$} \psi)^2$ is spin-dependent. Fock exchange interaction gives another spin-dependent interaction; e.g. $G/8N_c(\bar\psi\sigma^{\mu\nu} \psi)^2$ is generated through the Fierz transformation. Thus we can discuss competition or coexistence of chiral transition and magnetic transitions in a consistent way. This work is now in progress.

Instead, we briefly discuss here the relation between chiral symmetry and spin density wave. Utilizing the underlying chiral symmetry ($SU(2)_L\times SU(2)_R\simeq O(4)$) we assume non-vanishing pseudosclar condensate as well as scalar one in quark matter:
\begin{eqnarray}
\langle\bar{\psi}\psi\rangle&=&\Delta\cos\theta (\bf r)\nonumber\\
\langle\bar{\psi}i\gamma_5\mbox{\boldmath$\tau$}\psi\rangle&=&{\hat \bpi} ({\bf r})\Delta\sin\theta(\bf r),
\end{eqnarray}
where ${\hat \bpi}$ is a unit vector in the isospin space, and the phase angle $\theta$ and ${\hat \bpi}$ should be spatially dependent. Taking the simplest ansatz, $\theta({\bf r})={\bf q}\cdot{\bf r}, ~{\hat\bpi}_3=1$, we call the configuration dual {\it chiral density wave} (DCDW)  sketched in Fig.~5. We shall see that the amplitude $\Delta$ provides an effective mass for quarks and the phase degree of freedom $\theta={\bf q}\cdot{\bf r}$ gives rise to a magnetic property.
\begin{figure}[h]
\begin{minipage}{0.4\textwidth}
\begin{center}
\includegraphics[width=4.8cm]{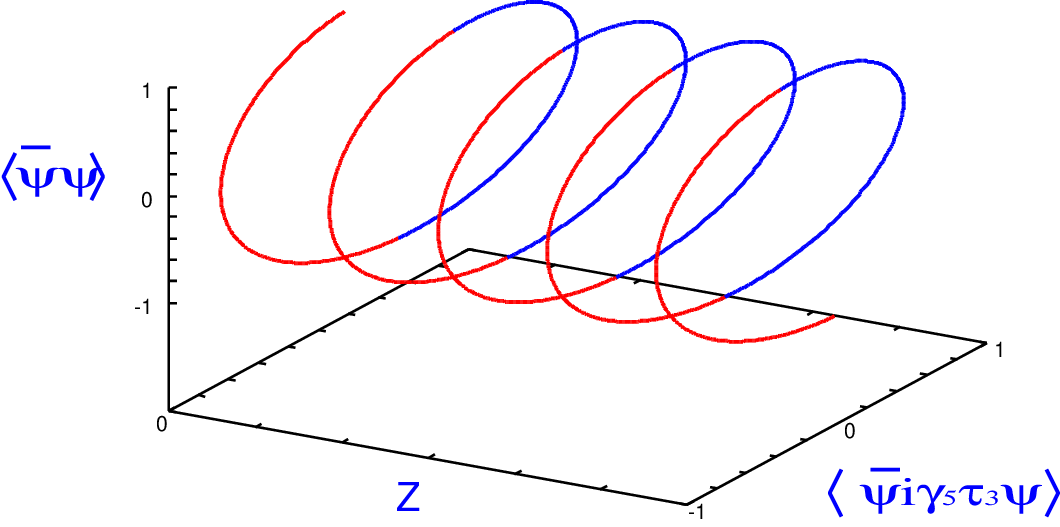}
\end{center}
\caption{Dual chiral density wave in the chiral space.}
\label{Fig:diagram}
\end{minipage}
\hspace{\fill}
\begin{minipage}{0.56\textwidth}
\vspace*{-0.4cm}
\begin{center}
\includegraphics[width=4.5cm]{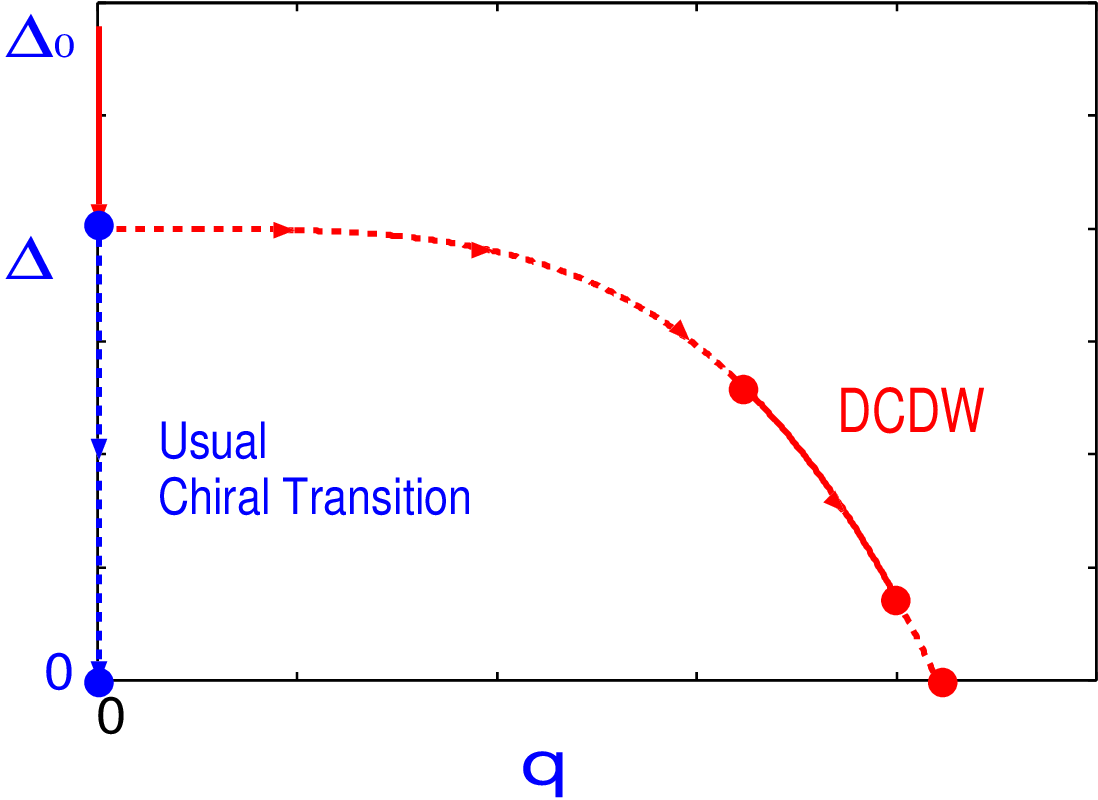}
\end{center}
\caption{Possible chiral restoration paths on the $\Delta-q$ plane. Dotted lines show the first order phase transitions.}
\label{path}
\end{minipage}
\end{figure}
Both condensates construct the complex order parameter of the chiral transition in this case in contrast with the scalar condensate in the usual discussions (Fig.~6)
\footnote{ Similar ideas of chiral density wave have been proposed as relevant phases in the large $N_c$ limit \cite{chi}, where only non-uniform scalar condensate has been considered. Thus they use only $Z_2$ symmetry, while we utilize $O(4)$ symmetry.}
.

Some results are presented in Figs.~7,8 by using the NJL model in the chiral 
limit \cite{dcdw}. 
Different from the usual result denoted by the thin dotted curve, DCDW appears around the critical density with finite momentum $q$ (Fig.~7). Restoration of chiral symmetry is then delayed by the appearance of DCDW. The magnitude of $q$ is $O(2k_F)$, which suggests that the {\it nesting} effect of the Fermi surface should be responsible to emergence of DCDW \cite{ove}. Actually we can show that the pseudoscalar correlation function  diverges at finite $q$ with $O(2k_F)$ at the critical density \cite{dcdw}.
\begin{figure}[h]
\begin{minipage}{0.48\textwidth}
\begin{center}
\includegraphics[width=5cm]{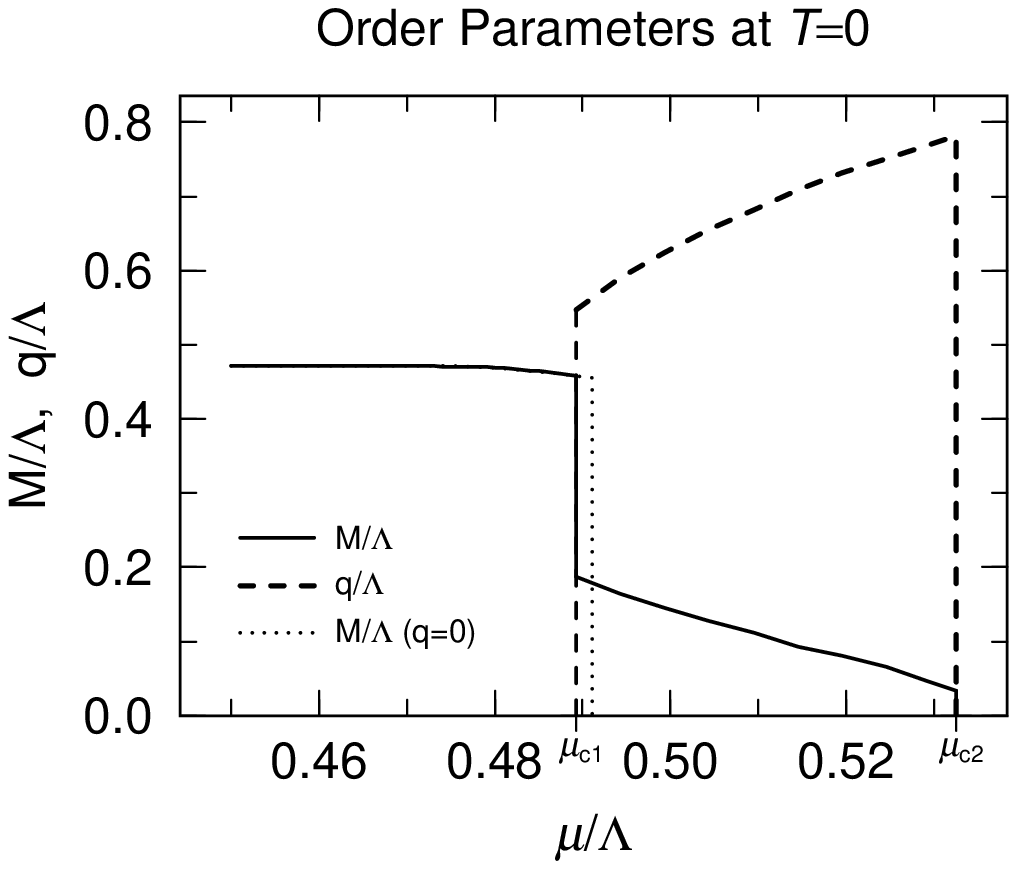}
\end{center}
\caption{Wave number $q$ and the dynamical mass $M=-2G\Delta$ are plotted at 
$T=0$. 
Solid (dotted) line for $M$ with (without) DCDW, 
and dashed line for $q$.}
\label{op1}
\end{minipage}
\hspace{\fill}
\begin{minipage}{0.48\textwidth}
\vspace*{-0.4cm}
\begin{center}
\includegraphics[width=5cm]{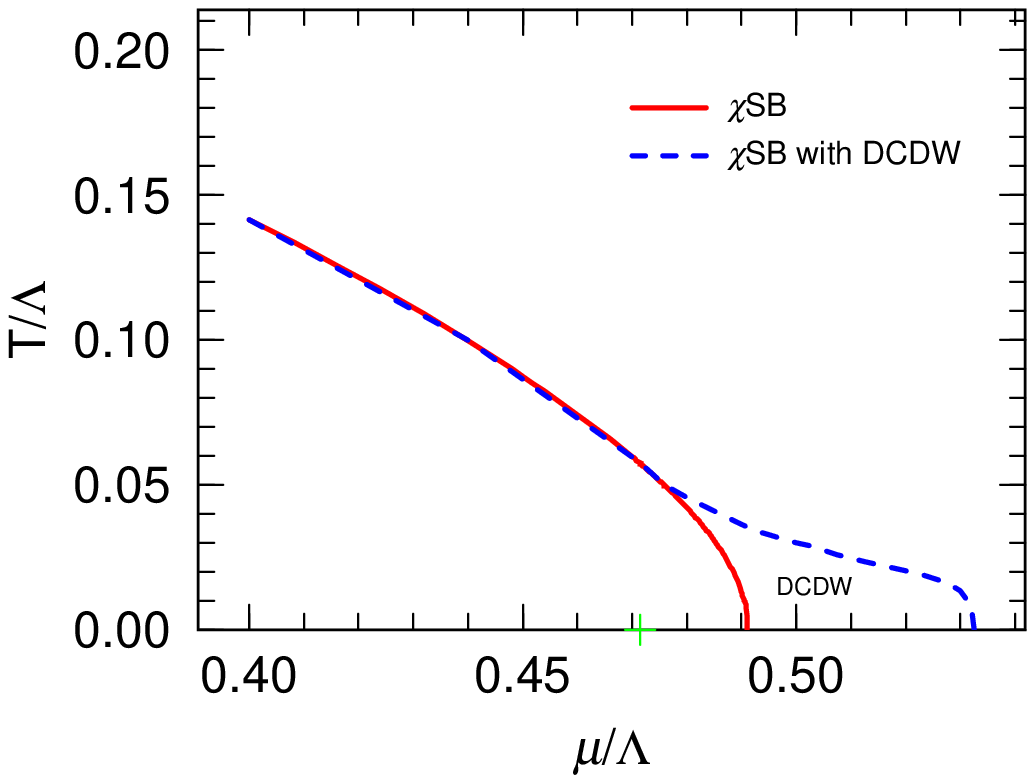}
\end{center}
\caption{Phase diagram of chiral transition on the temperature-density plane. DCDW appears at rather low temperature.}
\label{phaseT}
\end{minipage}
\end{figure}
Axial-vector current is then vanished, but 
magnetization of quark matter is generated instead,    
$
\langle \bar\psi\Sigma_z \psi\rangle=M_z\cos({\bf q}\cdot{\bf r}),
$
which implies there develops SDW in the DCDW phase.

It should be interesting to see a similarity of the configuration with pion condensation in hadronic matter, where nucleons take an anti-ferromagnetic spin ordering in the presence of classical pion field \cite{pio}. It suggests a kind of hadron-quark continuity.
\section{Summary and concluding remarks}

We have discussed some magnetic aspects of QCD on the temperature-density plane. For the ferromagnetic transition, we have studied the static magnetic susceptibility of quark matter within Fermi-liquid theory. We have seen the important roles of static and dynamic screenings: static screening gives $g^4\ln g^2$ term at $T=0$, while dynamic screening $T^2\ln T$ term at $T\neq 0$, as leading order contributions. The latter is a novel non-Fermi liquid effect. 

For more realistic description of magnetic aspects of QCD at moderate densities, we must take into account some non-perturbative effects. In particular their relation with chiral symmetry should be important. We must consider magnetic transitions and chiral transition in a consistent way. Although this program has not been completed yet, we have seen, for example, appearance of spin density wave near the phase boundary of chiral transition by using NJL model.

In recent papers Nickel discussed the appearance of the non-uniform phase to show that the real kink crystal (RKC) phase may appear and the first-order chiral-transition line is replaced by the two second-order phase transition lines \cite{nic}. The tricritical point is then the Lifshitz point in this case. This possibility is interesting but more studies are needed to elucidate the relation between RKC and DCDW phases, while he suggested that RKC is more favored than DCDW phase. 

Recent studies using gauge/gravity correspondence may also shed light on magnetic aspects of QCD at moderate densities \cite{ber}.

\section*{Acknowledgements}

This work was partially supported by the 
Grant-in-Aid for the Global COE Program 
``The Next Generation of Physics, Spun from Universality and Emergence''
from the Ministry of Education, Culture, Sports, Science and Technology
(MEXT) of Japan  and the 
Grant-in-Aid for Scientific Research (C) (16540246, 20540267).


%


\medskip
\begin{thebibliography}{9}
  
\bibitem{wam} Braun-Munzinger P and Wambach J 2009 {\it Rev. Mod. Phys.} {\bf 81} 1031.

\bibitem{pio} Takatsuka T et al. 1978 {\it Prog. Theor. Phys.} {\bf 59} 1933.\\
             
\hskip -0.33cm       Tatsumi T, 1980  {\it Prog. Theor. Phys.} {\bf 63} 1252.\\
\hskip -0.33cm                     Akmal A and Pandharipande V R 1997 {\it Phys. Rev.} {\bf C79} 2261.
\bibitem{dcdw} Tatsumi T and Nakano E 2004  Dual chiral density wave in quark matter Preprint hep-ph/0408294.\\
\hskip -0.33cm                Nakano E and Tatsumi T 2005 {\it Phys. Rev.} {\bf D71} 114006.@

\bibitem{tat00} Tatsumi T 2000 {\it Phys. lett.} {\bf B489} 280.\\
\hskip -0.33cm                 Tatsumi T, Nakano E and Nawa K 2006 {\it Dark
	Matter}, p.39 (New York: Nova Science Pub.).

\bibitem{tat082} Tatsumi T and Sato K 2008 {\it Phys. Lett.} {\bf B663} 322.

\bibitem{tat083} Tatsumi T and Sato K 2009 {\it Phys. Lett.} {\bf B672} 132.\\
\hskip -0.33cm                  Sato K and Tatsumi T 2009 {\it Nucl. Phys.} {\bf A826} 74.

\bibitem{mag} Woods P W and Thompson C 2006  
	{\it Compact stellar X-ray sources} 547.\\
\hskip -0.33cm 
Harding A K and Lai D 2006 {\it Rep. Prog. Phys.} {\bf 69}
	2631.





\bibitem{blo} Herring C 1966 {\it Magnetism IV} (New York: Academic press)


\bibitem{bru57} Brueckner K A and Sawada K 1957 {\it Phys. Rev.} {\bf 112}
               328.\\
\hskip -0.33cm          Shastray B S 1977 {\it Phys. Rev. Lett.} {\bf 38} 449.

\bibitem{man} Manuel C and Le Bellac 1997 {\it Phys. Rev.} {\bf D55} 3215.\\
 \hskip -0.19cm             Manuel C 2000 {\it Phys. Rev.} {\bf D62} 076009.

\bibitem{smi} Smith R P et al. 2008 {\it Nature} {\bf 455} 1220.

\bibitem{sch} Sch\"afer T and Schwenzer K 2004 Phys. Rev. {\bf D70} 054007;              114037.

\bibitem{deu} Deur A, Burkert V, Chen J P, Korsch W 2008 {\it Phys. Lett.} {\bf B665} 349.

\bibitem{hol} Holstein T, Norton R E and Pincus P 1973 {\it Phys. Rev.} {\bf
	B8} 2649.\\
\hskip -0.19cm               Ipp A, Gerhold A and Rebhan A 2004 {\it Phys. Rev.} {\bf D69}
	011901.

\bibitem{son} Son D T 1999 {\it Phys. Rev.} {\bf D59} 094019.

\bibitem{chi} Deryagin D V, Grigoriev D Yu and Rubakov V A 1992  
              {\it Int. J. Mod. Phys.} {\bf A7} 659.\\
\hskip -0.19cm               Shuster E and Son D T 2000 {\it Nucl. Phys.} {\bf B}573 434.\\ 
\hskip -0.19cm               Park B-Y, Rho M, Wirzba A and Zahed I 2000 {\it Phys. Rev.} {\bf D62} 034015.\\
\hskip -0.19cm               Rapp R, Shuryak E and Zahed I 2001 {\it Phys. Rev.} {\bf D63} 034008. 

\bibitem{ove} Overhauser A W,  Phys. Rev. Lett.{\bf 4} (1960) 462;   
                  Phys. Rev. {\bf 128} (1962) 1437. 

\bibitem{nic} Nickel D 2009 {\it Phys. Rev. Lett.} {\bf 103} 072301; {\it Phys. Rev.} {\bf D80} 074025.

\bibitem{ber} Gorsky A and Krikun A 2009 {\it Phys. Rev.} {\bf D79} 086015.\\
\hskip -0.19cm               Bergman O, Lifschytz G, Lippert M, 2009 {\it Phys. Rev.} {\bf D79} 105024.
              

\end{thebibliography}
\end{document}